# Cation nonstoichiometry and its impact on nucleation, structure and defect formation in complex oxide heteroepitaxy – LaCrO$_3$ on SrTiO$_3$(001)

L. Qiao[a], K.H.L. Zhang[a], M.E. Bowden[b], V. Shutthanandan[b], R. Colby[b], Y. Du[b], B. Kabius[b], P.V. Sushko[c], S.A. Chambers[a]

[a]*Fundamental and Computational Sciences Directorate*

*Pacific Northwest National Laboratory, Richland, WA, USA*

[b]*Environmental Molecular Sciences Laboratory*

*Pacific Northwest National Laboratory, Richland, WA, USA*

[c]*Department of Physics and Astronomy and London Centre for Nanotechnology,*

*University College London, Gower Street, London, WC1E 6BT, United Kingdom*

## Abstract

Our ability to design and fabricate electronic devices with reproducible properties using complex oxides is critically dependent on our ability to controllably synthesize these materials in thin-film form. Structure-property relationships are intimately tied to film and interface composition. Here we report on the effects of cation stoichiometry in LaCrO$_3$ heteroepitaxial films prepared using molecular beam epitaxy. We show that LaCrO$_3$ films grow pseudomorphically on SrTiO$_3$(001) over a wide range of La-to-Cr atom ratios. However, the growth mode and structural quality are sensitive to the La-to-Cr ratio, with La-rich films being of considerably lower structural quality than Cr-rich films. Cation mixing occurs at the interface for all La-to-Cr ratios investigated, and is not quenched by deposition at ambient temperature. Indiffused La atoms occupy Sr sites in the substrate. The presence of defects in the SrTiO$_3$ substrate is implicated in promoting La indiffusion by comparing the properties of LaCrO$_3$/SrTiO$_3$ with those of LaCrO$_3$/Si, both prepared at ambient temperature. Additionally, pulsed laser deposition is shown to result in more extensive interfacial mixing than molecular beam epitaxy for deposition at ambient temperature on Si.

## 1. Introduction

Since the discovery of high-temperature superconductivity in cuprates at the end of the last century, complex oxide films and multilayers have been of significant interest in condensed-matter physics and materials science, as well as electronic device technology, because of their wide range of physical properties. The interactions of multiple intrinsic order parameters (charge, spin, orbital, and lattice), combined with extrinsic properties (e.g. reduced dimensionality and thermal and photon activation) have produced a rather complex phase space of properties. In this space, a broad range of functionalities can be present including superconductivity, photoconductivity, photovoltaic effects, ferromagnetism, ferroelectricity and multiferroic behavior. The rich range of optical, dielectric, electronic and magnetic behavior in these materials, along with their structural, chemical and thermal stabilities endow these materials with potential new functionalities not possible in conventional semiconductors.

Reproducible and reliable oxide electronic devices cannot be fabricated and the mechanisms for their operation cannot be understood unless control over epitaxial film growth and heterojunction formation is achieved. Historically, the development of robust and controllable epitaxial film growth methods has led to the maturation of other electronic materials technologies. For example, the perfecting of III-V semiconductor heteroepitaxy by molecular beam epitaxy (MBE) and metal organic chemical vapor deposition (MOCVD) in the 1980s and 1990s has led to the invention and development of a number of novel device technologies based on these materials. Complex oxides are now viewed as one of the next frontiers in electronic materials. Pulsed laser deposition (PLD) is a very popular deposition method for complex oxide films because of its relatively low cost and ease of use. MBE has been less commonly used for complex oxides because it is generally more expensive and more difficult to use. A significant number of publications claiming exotic and novel electronic phenomena have appeared based mostly on PLD-grown material. However, the rigor in film growth and materials characterization associated with the development of III-V based materials has in general not been present in much of the complex oxide activity. A good example is the discovery of conductivity at the interface of polar and nonpolar insulating perovskites, such as $LaAlO_3/SrTiO_3$ (LAO/STO).[1-4] Explanations of this unexpected behavior range from classic electronic reconstruction,[5-7] based on the assumptions of fully stoichiometric films and atomically perfect interfaces, to oxygen vacancy formation[8-10] and unintentional doping by cation intermixing.[11-14] However, despite several years of work along with hundreds of publications and conference presentations, consensus is still lacking, quite possibly as a result of significant sample variation from lab to lab due to a lack of control over film and heterojunction growth and inadequate materials characterization. The question of stoichiometry in complex oxide films and its relationship to functional properties is largely unexplored. In PLD growth, it is often tacitly assumed that composition is preserved in going from the ablation target to the film. However, this assumption may not be valid, particularly when the substrate is exposed to only one part of the plume.[15-17] Evidence for changes in the cation atom ratios during laser ablation has recently been generated for perovskites such as STO,[18, 19] $BaTiO_3$,[20] $PbTiO_3$, $LaAlO_3$,[21] $SrRuO_3$,[22] $LaMnO_3$,[23] and $SrMnO_3$.[24] Moreover, slight deviations in composition have proven to induce dramatic changes in the structure and properties of some complex oxides.[25, 26] Composition control in PLD is generally thought to be achievable by tuning the laser energy density to exceed the threshold ablation energy for the various cations in the target.[27, 28] However, due to the complexity of the interaction between incoming laser photons and the target, our knowledge about the physics of the ablation process is limited.

In comparison, composition control is more straightforward in MBE, at least in principle. Because each metal source is independently controlled, one can tune film composition by tuning metal evaporation rates and accounting for the dependences of sticking coefficients at the growth temperature. Moreover, MBE has the potential to produce films of higher purity and lower defect density than PLD owing to the higher purity of commercially available metals compared to oxides, and the lower characteristic energies of atomic species emanating from effusion cells compared to those associated with ablation targets. The downside of oxide MBE is its higher cost and greater level of difficulty. Films are often grown under oxygen-rich conditions using a source of activated $O_2$.[29] As a result, precise composition control remains a major challenge, especially when the different cations exhibit large differences in oxygen affinity. Extents of surface oxidation of the hot metal charges in effusion cells vary considerably across the periodic chart, and can have a profound effect on evaporation and sublimation rates. *In-situ* flux monitoring techniques such as quartz crystal microbalance (QCO) and atomic absorption technology,[30, 31] as well as indirect methods such as reflection high-energy electron diffraction (RHEED),[32, 33] have been used to monitor fluxes and film stoichiometry for some complex oxides. However, the sensitivity and reproducibility for each of these monitoring techniques are material dependent. None of these have enough sensitivity for all elements typically found in complex oxides to allow composition control down to the atomic percent level, at least thus far.

Here we report on the effect of composition variation for $LaCrO_3$ (LCO) heteroepitaxy on STO(001) by MBE, examining film nucleation, structure, morphology, interfacial mixing and defect generation. We have previously demonstrated that stoichiometric LCO grows coherently on STO(001) up to a thickness of at least 900 Å and that polar LCO/nonpolar STO interfaces are insulating.[34, 35] The relatively small in-plane lattice mismatch (~0.5% for pseudocubic LCO) facilitates the heteroepitaxy of LCO on STO(001), and allows the effect of film composition on structure to be systematically investigated. We demonstrate that crystalline quality, film surface morphology and defect density are sensitive to the La-to-Cr cation ratio. We also discuss the role of substrate quality and deposition method in promoting interfacial mixing.

## 2. Results and Discussion

### 2.1. Film growth and *in-situ* characterization

A set of RHEED specular beam intensity oscillations as a function of growth time are shown in **Fig. 1** for 500Å-thick Cr-rich (La:Cr = $x$ = 0.85), stoichiometric ($x$ = 1.00) and La-rich ($x$ = 1.10) LCO films. For the stoichiometric film, the RHEED oscillations persist with uniform amplitude throughout the entire film growth, indicating a stable and persistent two-dimensional (2D) layer-by-layer growth. Persistent RHEED oscillations were also observed for the Cr-rich film, although a low-frequency modulation is superimposed on the oscillations. In contrast, the oscillation amplitude decays to near zero for the La-rich film, indicating the cessation of 2D layer-by-layer growth.

We also show in Fig. 1 RHEED patterns along both [110] and [100] crystallographic directions and AFM images for an HF-etched and tube furnace annealed STO(001) substrate, and 500 Å films with different compositions. The surfaces of both the stoichiometric and Cr-rich films retain the step-terrace structure of the substrate with a minimum step height of ~4 Å, consistent with the persistent RHEED oscillations. In contrast, the La-rich film surface roughens considerably, obliterating the terrace-step structure of the substrate, consistent with the decay of the RHEED intensity oscillations. The RHEED patterns for both stoichiometric and Cr-rich films are strong and streaky, indicative of well-ordered and flat surfaces. Additionally, both exhibit a (2×2) surface reconstruction. However, the surface of La-rich film exhibits a much weaker (1×1) pattern, revealing considerably more disorder associated with the rough surface seen in AFM.

Core-level and valence band (VB) spectra are shown in **Fig. 2** for a series of films with $x$ ranging from 0.92 to 1.25. A low-energy electron flood gun had to be used to compensate the positive surface charge accompanying photoemission when the film surfaces were found to be insulating. In order to directly compare binding energies, the O $1s$ peaks (not shown) were all shifted to 530.0 eV. For each specimen, the same shift was then applied to La $4d$, Cr $2p$ and VB spectra. The La $4d$ and Cr $2p$ binding energies and line shapes are characteristic of $La^{3+}$ and $Cr^{3+}$ for all cation ratios, and change little with $x$. The VB spectra for x ranging from 0.92 to 1.11 changes in ways that are largely predictable based on the known electronic structure of LCO. The feature at ~ 2 eV, dominated by Cr $3d$ $t_{2g}$ with an admixture of O $2p$ character,[34, 35] drops in intensity relative to the rest of the VB with increasing $x$ as the Cr fraction in the film drops. However, the overall shape of the VB remains characteristic of that for LCO for $x$ = 0.92 to 1.11, and there is no evidence for secondary phases in these data. In contrast, the portion of the VB from ~3 to ~9 eV undergoes a substantial change in shape for $x$ = 1.25. The valley fills in and the shoulder at ~8 eV is no longer resolved. Even though the RHEED patterns (not shown) remain characteristic of those of LCO, this portion of the VB becomes very similar in appearance to that of La$_2$O$_3$, which is also shown in Fig. 2, right-hand panel. The latter spectrum was measured from an La$_2$O$_3$ reference film, prepared by depositing La metal in excess O$_2$ on Si(001) at ambient temperature. For these data, we conclude at $x = 1.25$, at least some of the excess La is excluded from the lattice and forms a disordered secondary phase of La$_2$O$_3$.

## 2.2. Film crystallography

XRD data reveal that representative films with $x$ values ranging from 0.85 to 1.10 contain epitaxial LCO as the majority phase, and do not reveal the presence of any detectable secondary phases. We show in **Fig. 3a** a representative {110} $\varphi$ scan for a stoichiometric LCO film and the STO substrate. The presence of four-fold symmetry, together with overlapping film and substrate Bragg peaks, reveal the expected cube-on-cube structure with the epitaxial relationship (001)$_{LCO}$ $\parallel$ (001)$_{STO}$ and [110]$_{LCO}$ $\parallel$ [110]$_{STO}$. The $\varphi$ scans for the other films are the same.

**Fig. 3b** shows a set of $\theta$-$2\theta$ scans near the (002) Bragg peak for four approximately stoichiometric films with different film thicknesses: 100 Å ($x = 1.00$), 200 Å ($x = 0.97$), 500 Å ($x = 1.00$) and 900 Å ($x = 0.94$). All four films exhibit finite thickness fringes with periods that scale with film thickness. The Bragg peak for the 100 Å film peak is rather weak and very broad because of the small thickness, but the thickness fringes are nevertheless clearly resolved. Thickness fringes are seen in films with uniform thickness and out-of-plane lattice constant. Thus, the presence of these fringes in all films reveals good crystallinity. We also performed (002) rocking-curve measurements for 200, 500 and 900 Å films (not shown). All films exhibit the same rocking-curve shape and full-width at half-maximum (FWHM) value as their respective STO substrates, thus indicating that film quality is limited by substrate quality. Typical FWHM values range from 0.008° to 0.02°. Both out-of-plane and in-plane lattice parameters were calculated by simultaneously fitting the (00$l$) ($l$ = 1, 2, 3), (113), (123) and (114) peaks. The resulting lattice parameters for all four films plus those for a stoichiometric 15 u.c. LCO/STO (as determined by RBS channeling) are plotted in **Fig. 3c**. All films are coherently strained to the substrate, and the $c$ lattice parameter depends on cation ratio, as seen in the inset.

In **Fig. 3d** we compare the structural quality and out-of-plane lattice parameters of stoichiometric and nonstoichiometric films. The Cr-rich film exhibits oscillations slightly weaker than those of the stoichiometric film, indicating slightly reduced crystal quality. In contrast, the La-rich film shows almost no oscillation amplitude, revealing poorer ordering and/or a rougher surface. Rocking-curve measurements confirm that the Cr-rich and stoichiometric films are of comparable quality since their (002) rocking-curve widths are the same as those of their substrates, whereas the La-rich film exhibits a larger width than its substrate (typically larger than 0.1°). As seen in the inset to Fig. 3c, the three films portrayed in Fig. 3d have the same $a$ value as STO, revealing a critical thickness in excess of 500 Å, independent of stoichiometry from $x$ = 0.85 to 1.10. Point defects associated with nonstoichiometry may provide strain relief in lieu of line defects, thereby enabling a larger critical thickness than nominally expected for 0.5% in-plane lattice mismatch.

We also see from Fig. 3d that the Bragg angle for these films decreases with increasing $x$, revealing a concomitant increase in the out-of-plane lattice parameter, $c$ (**Fig. 3c**). In order to gain some understanding of this trend, we carried out first-principles calculations to determine the effect of various kinds of defects on the value of $c$ for LCO that is coherently strained to the substrate. We focus on the Cr and La defects only: (i) anti-site defects in which a La atom occupies a Cr site (La$_{Cr}$) and a Cr atom occupies a La site (Cr$_{La}$), (ii) neutral La and Cr vacancies, and (iii) neutral La and Cr interstitials. The anti-site defects do not require charge compensation since Cr and La have the same oxidation state, while the vacancies and interstitials produce three holes and three extra electrons, respectively. Here we considered one defect per $2\sqrt{2}a \times 2\sqrt{2}a \times 4c$ supercell, which corresponds to $x = 0.94$ for $Cr_{La}$, $x = 0.97$ for Cr vacancy and La interstitial, $x = 1.03$ for La vacancy and Cr interstitial, and $x = 1.06$ for $La_{Cr}$. Substitutional Cr in $Cr_{La}$ displaces from the La site and gives rise to the axial and equatorial configurations, as expected for strained epitaxial LCO. Hence, the formation energies of $Cr_{La}$ and the values of $c$ were averaged with the corresponding Boltzmann factors for T = 300 K. Interstitial Cr atom is located in the $BO_2$ planes that are either parallel or perpendicular to the plane of the interface. In our case, the difference of the total energies calculated for these configurations is significant enough that only the former can be considered. Similarly, only the site in the AO plane parallel to the plane of the interface need to be considered for the interstitial La defect.

The phase diagram for all considered defects is shown in **Fig. 4**. The dashed horizontal line indicates the oxygen chemical potential calculated taking into account the background $O_2$ pressure and substrate temperature during the LCO film deposition.[36] These calculations reveal that the ideal LCO, La and Cr vacancies, and both anti-site defects are feasible; their relative contributions can be controlled by the La:Cr ratio. Interstitial Cr and La defects can be disregarded because their formation requires much lower oxygen pressures than what was used in this work.

The effect of the defects on the lattice parameter $c$ is also shown in **Fig. 4**. The optimal $c$ for the defect-free epitaxial LCO is 3.820 Å. For the considered concentration of the La and Cr vacancies (~3%), the value of $c$ decreases by 19 and 8 mÅ, respectively. The value of $c$ increases by 34 mÅ in the case of $La_{Cr}$ and decreases by 6 mÅ in the case of $Cr_{La}$. Since the La and Cr vacancies are stable in a relatively narrow range of the La:Cr ratio only, we conclude that the observed expansion of both $c$ and the unit cell volume with increasing $x$ (Fig. 3c) can be ascribed to the anti-site defects. The dependence of $c$ on $x$ can be rationalized if we consider the ionic radii of $La^{3+}$ (1.03 Å) and $Cr^{3+}$ (0.62 Å). [37] However, the non-symmetric change of $c$ (+0.034 Å for $La_{Cr}$ and –0.006 Å for $Cr_{La}$) suggests that the ion-size effect is only one of several factors contributing to this dependence, such as co-existence of several types of defects. For example, if the $La_{Cr}$ defects and the Cr vacancies are present in comparable concentrations in the La-rich LCO, their average effect on the value of $c$ will be $+34 - 19 = 15$ mÅ, which is close to what is observed experimentally. In addition, cation defects modify the magnetic structure of LCO. Indeed, $La_{Cr}$ can be considered as a vacant spin site in the G-type antiferromagnetic network of LCO, while in the case of $Cr_{La}$, the substitutional Cr couples ferromagnetically with half of the nearest lattice Cr ions and antiferromagnetically with the other half. Such perturbations of the magnetic order result in larger interatomic distances locally and contribute to the increase of $c$. These magnetic defects, together with the ion size effect as such, introduce disorder into the film, thus decreasing film quality, as observed.

The coherent in-plane strain is also revealed by the (103) reciprocal space maps (RSMs) for a series of stoichiometric films in **Fig. 3e**. There is no change in the out-of-plane momentum transfer between substrate and film for any of these samples. From the RSMs, it is also clear that film and substrate peaks have the same FWHM along H, revealing comparable crystal quality. The larger film FWHM along L results from finite-size broadening. The critical thickness for near stoichiometric film exceeds 900 Å, which is at least five times larger than the theoretical predication (183 Å) based on an elastic theory model.[35]

### 2.3. Local structure

Crystallographic quality and the local structural environments about La and Cr sites were also determined by RBS in channeling mode and rocking-curve measurements. **Fig. 5a** shows the RBS spectra in both channeling and random directions for films with different La-to-Cr atom ratios. The channeling spectra for both stoichiometric and Cr-rich films show low yield in the La surface peak region and a small extent of dechanneling. In contrast, the channeling spectrum for the La-rich film exhibits high counts in the La peak and more extensive dechanneling. A small energy window to the left of the La surface peak was used to determine the minimum yield ($\chi_{min}$), which is the ratio of the yield in the channeling geometry to that in a random direction. The $\chi_{min}$ values are 1.21%, and 0.64% for the stoichiometric and Cr-rich films, respectively, revealing excellent crystalline quality. In contrast, $\chi_{min}$ is 8.36% for the La-rich film, revealing substantial disorder, consistent with the XRD results.

Additional insight into the structural quality of these specimens can be gained from RBS rocking curve measurements. We show in **Fig. 5b** an RBS spectrum indicating the La regions of interest (LCO film surface, LCO/STO interface, and La diffusion into the STO substrate) and in **Fig. 5**c ~ **e** we plot the La angular scans along [001] for three different La-to-Cr atom ratios. The yields are integrated over the same depth regions for each of the films, and have been normalized by the values obtained in the random direction. Fig. 5c shows RBS angular scans of the three La-containing regions for the stoichiometric sample, along with Sr signal from STO(001). The intensity minimum for the near-interface La is as low as that for Sr, revealing that La is as well ordered in the A sites as is Sr. In contrast, the La "surface" yield along the channeling direction doesn't drop as much as that for both Sr and near-interface La, indicating disorder in the near-surface region, possibly due to surface oxidation to $LaCrO_4$ due to air exposure.[34, 35] There is no evidence for interstitial La near the interface. Interstitial La would result in a small peak at the bottom of the rocking curve for indiffused La. Additionally, indiffused La exhibits a rocking curve similar to that of Sr, indicating that indiffused La atoms are predominantly at Sr sites in the STO.

The critical angle ($\psi_c$) for channeling is also an important factor in RBS rocking-curve measurements since it is directly related to atomic displacements. Based on Lindhard's continuum mode,[38-42] the measured half-width ($\psi_{1/2}$) of an angular scan depends on two characteristic distances, the minimum distance for channeling ($r_{min}$) and the Thomas-Fermi screening length ($a$). The relevant formula is,

$$\psi_{1/2} = \psi_c = \psi_1 \left[ \frac{1}{2} \ln \left( \frac{3a^2}{r_{min}^2} + 1 \right) \right]^{1/2} \qquad (1)$$

where $r_{min}$ is the estimated minimum distance an ion travels to remain in the channeling direction in the continuum model, and is a function of the thermal vibration amplitude ($\rho$) and the projected atomic displacement ($u$), as viewed along channeling direction. Both $\psi_1$ and $a$ are functions of the target atomic number ($Z_2$) and the lattice spacing ($d$) along the channeling direction. The relevant formulas are $\psi_1 = (2Z_1 Z_2 e^2/Ed)^{1/2}$ and $a = 0.8853 a_0 (Z_1^{1/2} + Z_2^{1/2})^{-2/3}$, where $Z_1$ is the projectile atomic number, $E$ is the projectile energy, and $a_0$ is the Bohr radius. Thus, $\psi_{1/2}$ is directly dependent on the values of three factors: the target atomic number ($Z_2$), the atom displacement ($u$) and the lattice spacing ($d$). Thus, the observation of a larger $\psi_{1/2}$ for La (both near-surface and near-interface) than for Sr is ascribed to the larger $Z$ for La and the smaller lattice spacing along [001] direction for the LCO film than for bulk STO. However, the value of $\psi_{1/2}$ for indiffused La is lower than that for La in the bulk of the film, or for Sr in the STO. This result suggests that interdiffused La atoms are slightly laterally displaced as viewed along [001]. Calculations also indicate that while the displacements for Sr in the STO substrate and La away from the interface are ~0.01 and 0.02 Å, respectively, the estimated lateral displacement for indiffused La atoms within the STO is in excess of ~0.04 Å. This result is not unexpected considering that the interface is a quaternary due to site exchange for all four cations.

The rocking curves for the Cr-rich film are shown in Fig. 5d and exhibit nearly the same behavior as those for the stoichiometric film, indicating a high degree of crystal quality near the interface, slight disorder near film surface, and substitutional La at Sr sites within the mixed region. However, the structural quality of the La-rich film is lower than that of other two as revealed by the higher values of the intensity minima at $0°$ tilt angle, as seen Fig. 5e.

We have also measured rocking curves for Cr backscattering for these three films. Due to the much lower RBS sensitivity for Cr compared to La, the counting statistics are not as good as for La. Nevertheless, some qualitative similarities exist. As seen in **Fig. 5f**, the angular scan for the La-rich film exhibits a higher intensity at the minimum value than the other two films, revealing more structural disorder throughout the La-rich film, consistent with RBS and XRD. Therefore, disorder occurs not only at the La site, but also at the Cr site in the La-rich film.

## 2.4. Interdiffusion at the interface

**Fig. 6a** shows random RBS spectra for the three LCO films with different La:Cr ratios measured at a scattering angle of 150°. This geometry yields better mass resolution due to the shorter path-length for the backscattered beam and minimal inelastic-scattering-induced broadening. By performing least-squares fits to RBS simulations, we are able to obtain quantitative atomic percentages for both cations.[35] The RBS La:Cr ratios for the three films were 0.85, 1.00 and 1.10, as used throughout the text up to this point for these samples. We fit the spectra (using SIMNRA) with both abrupt and optimized intermixing models, as displayed in **Fig. 6a** and **b**. The abrupt model does not predict the nonzero counts in the valley between low-energy side of La peak and leading edge of Sr plateau. In order to predict the presence of counts in the valley, indiffusion of La and outdiffusion of Sr must be included in the model. The statistical noise associated with the low counts in the valley precludes obtaining a clear best fit to any one diffusion profile. However, the La atom profile which qualitatively accounts for the $x = 1.00$ spectrum is shown in **Fig. 6c**. This diffusion profile is modeled as discrete compositional steps in which the La concentration beings to drop before arriving at the interface (at a depth of ~500Å) and persists to a depth of more than 1000 Å, albeit at very low concentrations. Together with the rocking curve data shown in Fig. 5, these spectra show that A-site intermixing occurs to comparable extents in all three samples.

We now consider the roles of substrate temperature and sub-surface defect density in promoting intermixing. To determine the effect of temperature, we grew a 500Å stoichiometric, but amorphous LCO film on STO at ambient temperature. The amorphous character of the film was confirmed by RHEED and XRD. The RBS spectrum for this amorphous film is shown and compared to that for an analogous epitaxial film in **Fig. 7a**. However, the amorphous film exhibits an extent of intermixing similar to that of the epitaxial film, based on the nonzero counts in the valley between the La and Sr peaks (**Fig. 7b**). This result reveals that intermixing is not quenched simply by dropping the temperature. We hypothesize that defects in the STO substrate, possibly in the form of Sr vacancies resulting from the HF etch and tube furnace anneal required to generate a $TiO_2$-termianted surface, promote La indiffusion. To test this hypothesis, we grew an amorphous, stoichiometric LCO film on Si(001) at ambient temperature with the same atom fluxes. The RBS spectrum for this film is shown in Fig. 7a and b. In contrast to the LCO/STO interface prepared at room temperature, the LCO/Si interface is rather abrupt, as revealed by the absence of counts in the valley between the Cr and La peaks. This result supports our hypothesis, as electronic grade crystalline Si(001) is largely defect free.

We next explore the role of deposition technique in driving interfacial mixing. We compare LCO deposited at ambient temperature on Si(001) by MBE (**Fig. 7c** and **d**) with LaAlO$_3$ (LAO) deposited at ambient temperature on Si(001) by off-axis PLD at 10 mTorr O$_2$ (**Fig. 7e** and **f**). We did not deposit LCO by PLD because we were unsuccessful at fabricating a robust ablation target for LCO. In stark contrast to the rather abrupt LCO/Si interface prepared by MBE, the LAO/Si interface exhibits considerable La indiffusion, as revealed by the presence of counts in the valley between the La and Al peaks in Fig. 7f. This result is strongly suggestive of La implantation into Si as a result of energetic ions in the ablation plume, even at the relatively high O$_2$ pressure of 10 mTorr. The laser energy density on the single-crystal LAO target in the PLD system was ~3 J cm$^{-2}$, and this power level appears to give rise to ions in the plume that are sufficiently energetic to implant into crystalline Si.

## 2.5. Defects

Cross-sectional high- and low-angle annular dark field (HAADF/LAADF) STEM imaging provides corroborative insight into the structural quality and defect formation in both stoichiometric and nonstoichiometric films. HAADF-STEM is primarily sensitive to atomic number, but also somewhat sensitive to crystal quality (channeling), while LAADF-STEM is more sensitive to diffraction contrast and strain. For the case of nominally stoichiometric 500 Å LCO, HAADF-STEM intensity is fairly uniform for the first ~20 nm of the LCO, but decreasingly so near the surface, as shown in **Fig.8a** and **c**. The intensity is also lower near the surface, suggestive of either poorer crystal quality or a lower average Z (by volume), or both. (The two effects are frequently coupled.) As there was no demonstrable difference in the La:Cr ratio by EDS, and in light of the RBS results, the former explanation seems more likely. LAADF-STEM images in **Fig. 8b** and d support this argument, with strong and clear strain contrast visible again after the first ~20 nm of the film. Despite the apparent lines of strain contrast, there is no evidence for out-of-plane dislocations in STEM, more traditional dark-field TEM, or HRTEM. The additional LAADF-STEM contrast at the LCO/STO interface is typical of heteroepitaxially grown films. HAADF-STEM imaging also supports a diffuse interface.

In contrast, the 500 Å film for which $x = 1.10$ has a highly defective interface with frequent encapsulated pockets of amorphous material (**Fig. 9**). There is a more severe lateral inhomogeneity visible by HAADF-STEM (compare **Fig. 9a** and Fig. 8a). The encapsulated amorphous regions range in size from only a few unit cells (hardly differentiable from a dislocation core in cross-section) to tens of nm (**Fig. 9b**). The regions of the film above these defects are typically marked by defects and stacking faults. This result is consistent with AFM and RHEED observations. For the smaller inclusions, this typically resulted in a fairly coherent fault extending from the interface to the surface. Above the larger, roughly pyramid-shaped amorphous inclusions, the lattice planes bow upward, resulting in a lower-angle grain boundary. To more clearly demonstrate this result, a simple trick of laterally (along the [100] direction) compressing magnified images is employed for typical smaller (**Fig. 9c** and **e**) and larger (**Fig. 9d** and **f**) amorphous inclusions. EDS confirmed that the amorphous regions predominantly contain La and Cr, albeit with varying La:Cr ratios and, in some cases, a significant fraction of Sr.

### 3. Conclusion

We have brought several complementary analytical techniques, along with *ab initio* simulations, together to investigate the effect of cation nonstoichiometry on crystal quality, surface morphology, interfacial mixing, and defect generation in MBE-grown films of the polar perovskite oxide $LaCrO_3$ on $SrTiO_3(001)$. We find that this material is able to accommodate La-to-Cr ratios which differ from unity by as much as ~15% in the Cr rich direction without extensive loss of structural quality. The material buckles structurally more readily in the La-rich direction. In both directions, our data are consistent with anti-site defect formation as a way the LCO lattice accommodates La-to-Cr atom ratio mismatches. If we go far enough in the La-rich direction (i.e. to La:Cr = 1.25), excess La is expelled by the lattice into a disordered secondary phase ($La_2O_3$). Cation mixing occurs at the interface for all film compositions, and is not affected by nonstoichiometry in any obvious way. Comparison of data for LCO/STO(001) with that for LCO/Si(001), both prepared at ambient temperature, reveals that interfacial mixing is driven at least in part by defects in the STO. Likewise, comparison of MBE-grown LCO/Si(001) with PLD-grown $LaAlO_3$ (LAO)/Si(001), both prepared at ambient temperature, shows that the higher energies of atoms and ions in the laser ablation plume drive interfacial mixing much more than occurs in MBE, with its lower atom energies. These results provide guidelines and a framework to understand the best ways to prepare complex oxide heterojunctions for either fundamental research or novel device applications. In particular, this study demonstrates that the compositions of both $BO_2$ and AO sublattices of LCO can deviate significantly from the ideal stoichiometry without inducing major lattice structure changes. Given that low concentrations of defects can dramatically alter the electronic properties of perovskites, we conclude that structural analysis alone provides inadequate characterization of these materials and, in turn, can lead to misleading interpretation of the observed physical phenomena.

## 4. Experimental and Theoretical Details

As-received STO substrates were etched in buffered HF and annealed in air to produce $TiO_2$-terminated STO as described elsewhere [43]. Following sonication in acetone, isopropanol and distilled water, the substrates were loaded into a custom MBE chamber and exposed to activated oxygen from an electron cyclotron resonance (ECR) plasma at a chamber pressure of $2.0 \times 10^{-5}$ Torr for 30 minutes, resulting in no detectable carbon by x-ray photoelectron spectroscopy (XPS) [44, 45]. During LCO deposition, the oxidant was $O_2$ at a pressure of $\sim 6.0 \times 10^{-8}$ Torr, as determined by a residual gas analyzer (RGA), and the substrate temperature was $650 \pm 50^{\circ}C$, as measured by a two-color infrared pyrometer. Cr was deposited using an electron beam evaporator and the sublimation rate was monitored and controlled using *in situ* atomic absorption (AA) spectroscopy. La was evaporated using a high-temperature effusion cell and the La rate was measured before and after deposition with a QCO positioned at the substrate position. The Cr flux was adjusted through (AA) feedback control to achieve different La-to-Cr ratios (*x*). The typical growth rate was $\sim 40$ sec per unit cell (u.c.). The $O_2$ valve was closed immediately after growth and samples were cooled to room temperature in high vacuum to prevent oxidation of structural $Cr^{3+}$ in the near-surface region of LCO to $Cr^{5+}$ or $Cr^{6+}$ [34, 35].

Film growth and surface quality were monitored by *in-situ* RHEED at 15 keV. Generally, film growth was terminated when the RHEED specular beam intensity was at a maximum, if intensity oscillations were observed, to insure the deposition of an integer number of unit cells. Film thickness was determined by counting RHEED oscillations (if present), as well as from x-ray reflectivity (XRR) and Rutherford backscattering spectrometry (RBS). Film thicknesses ranged from 100 to 900 Å. Film composition was determined by *in-situ* high-energy-resolution XPS using a Scienta SES 200 electron energy analyzer with a monochromatic Al *Kα* x-ray source. La-to-Cr atom ratios were determined from La 4*d* and Cr 3*p* peak area ratios by calibrating with RBS and using SIMNRA simulations [46] on select samples. Local structure at specific sites was determined using RBS rocking curve measurements. Surface morphology was measured *ex situ* by tapping-mode AFM using a Digital Instruments Nanoscope III. Lattice parameters and overall crystal quality were determined using high-resolution x-ray diffraction (XRD) and reciprocal space mapping (RSM) with a Philips X'Pert diffractometer equipped with a Cu anode. A hybrid monochromator, consisting of four-bounce double crystal Ge (220) and a Cu x-ray mirror, was placed in the incident beam path to generate monochromatic Cu $K\alpha$ x-rays ($\lambda = 1.54056$ Å) with a beam divergence of 12 arc seconds. Crystal quality was determined by means of triple-axis rocking curve measurements in which an additional three-bounce Ge (220) channel-cut analyzer was placed in front of a proportional counter to obtain the same divergence as the incident beam. Select films were examined in cross-section using conventional and scanning transmission electron microscopy (TEM/STEM) with a probe-corrected FEI Titan microscope operated at 300 kV. X-ray energy dispersive spectra (EDS) were also collected in STEM at 300 kV with a 0.13 mrad EDAX detector. Cross-sections were prepared with an FEI Helios dual-beam FIB/SEM equipped with an Omniprobe manipulator using the lift-out technique and further thinned on a liquid nitrogen cooled stage with an Ar-ion source (0.5–2 kV), using a Fischione NanoMill.

Relative stability of the cation anti-site defects and neutral La and Cr vacancies and interstitials were calculated using a method developed for comparison of the free energies of surfaces with different stoichiometries. [36, 47, 48] In our case, this method allows one to determine preferential defect structures in the Cr-rich (La-poor) and Cr-poor (La-rich) conditions as a function of oxygen pressure and temperature. The free energies for solids have been approximated by the corresponding total energies calculated using the density functional theory (DFT); the density functional optimized for solids by Perdew *et al.* (PBEsol) [49] and the projected augmented waves method [50], as implemented in the Vienna *ab initio* simulation program (VASP) code [51] were used. The plane-wave basis set cutoff was set to 500 eV. All calculations were performed using the $2\sqrt{2}a_0 \times 2\sqrt{2}a_0 \times 4c$ supercell and the $\Gamma$ point of the Brillouin zone. The in-plane lattice constant was fixed at the value corresponding to that of the bulk SrTiO$_3$ ($a_0 = 3.905$ Å). The total energy was minimized with respect to the value of the lattice constant $c$ and internal coordinates of all atoms until the forces acting on atoms are below 0.01 eV/Å.

## Acknowledgements

This work was supported by the U.S. Department of Energy, Office of Science, Division of Materials Sciences and Engineering under Award #10122 (MBE growth and XPS), Division of Chemical Sciences under Award #48526 (XPS analysis and RBS measurements and analysis), and the EMSL William Wiley Postdoctoral Fellow program (TEM analysis). P.V.S. thanks the Royal Society for the support. The work was performed in the Environmental Molecular Sciences Laboratory, a national science user facility sponsored by the Department of Energy's Office of Biological and Environmental Research and located at Pacific Northwest National Laboratory.

**Figures and captions**

**Figure 1**. RHEED specular beam intensity oscillations as a function of growth time for 500 Å La-rich (La:Cr = $x$ = 1.10), stoichiometric ($x$ = 1), and Cr-rich ($x$ = 0.85) LCO films. Also shown are RHEED patterns along both [110] and [100] directions and AFM images for these films and a clean STO(001) substrate.

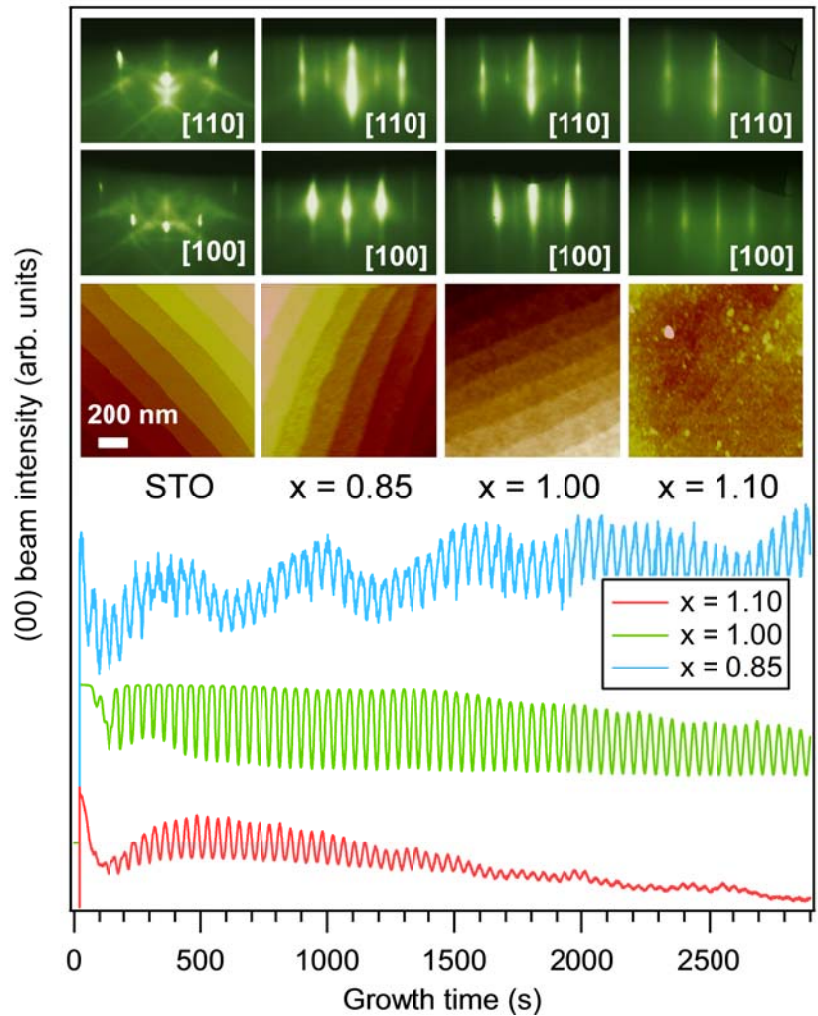

**Figure 2.** Core-level and valence band x-ray photoelectron spectra measured at normal emission for a series of LCO(001) films with *x* ranging from 0.92 to 1.25.

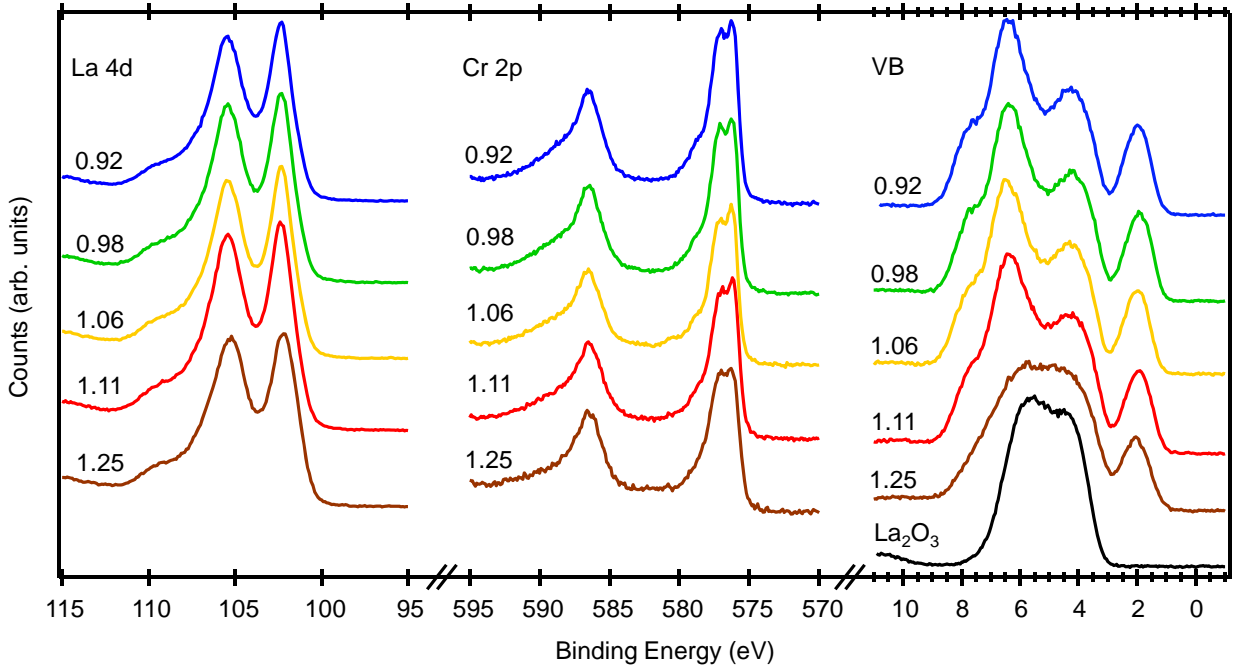

**Figure 3.** (a) Representative in-plane x-ray diffraction {110} *φ* scans for a stoichiometric LCO film and its STO(001) substrate. (b) High-resolution XRD patterns near the (002) Bragg peak for four near-stoichiometric LCO films with different thickness: 100 Å (*x* = 1.00), 200 Å (*x* = 0.97), 500 Å (*x* = 1.00) and 900 Å (*x* = 0.94). (c) Lattice parameters (both in-plane, open circles, and out-of-plane, closed circles) for a series of films as a function of film thickness. The dashed lines denote the lattice parameters for coherently strained thin (≤100Å) films of stoichiometric LCO.

The inset shows the lattice parameters as a function of film stoichiometry for three 500 Å LCO films with different $x$ values. (d) High-resolution XRD patterns near (002) Bragg peak for three 500 Å LCO films with different $x$ values. (e) (103) Reciprocal space maps for stoichiometric LCO films with different thicknesses. Both axes are referenced to the reciprocal lattice dimensions of STO (1 $r.l.u.$ = 2π/3.905 Å).

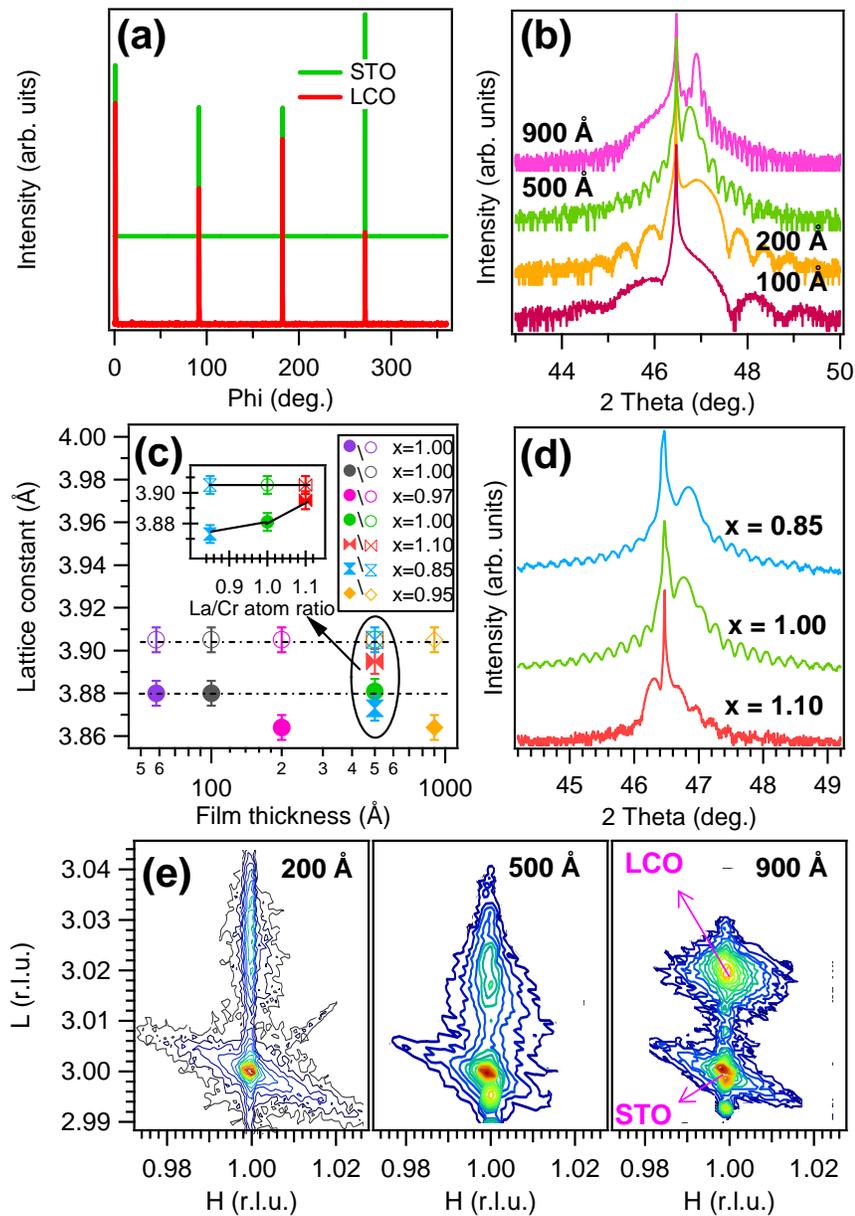

**Figure 4.** Phase diagram of the preferential cation defects in epitaxially strained LCO: 1 – non-defective LCO supported on STO(001), 2 – anti-site defect $La_{Cr}$; 3 – neutral Cr vacancy; 4 – neutral La vacancy; 5 – anti-site defect $Cr_{La}$; 6 – interstitial Cr, 7 – interstitial La. The numbers at the top of the figure show the changes of the *c* lattice parameter (in mÅ) for the considered La:Cr ratios. The horizontal dashed line indicates the oxygen chemical potential corresponding to the $O_2$ background pressure and the substrate temperature during the LCO film deposition.

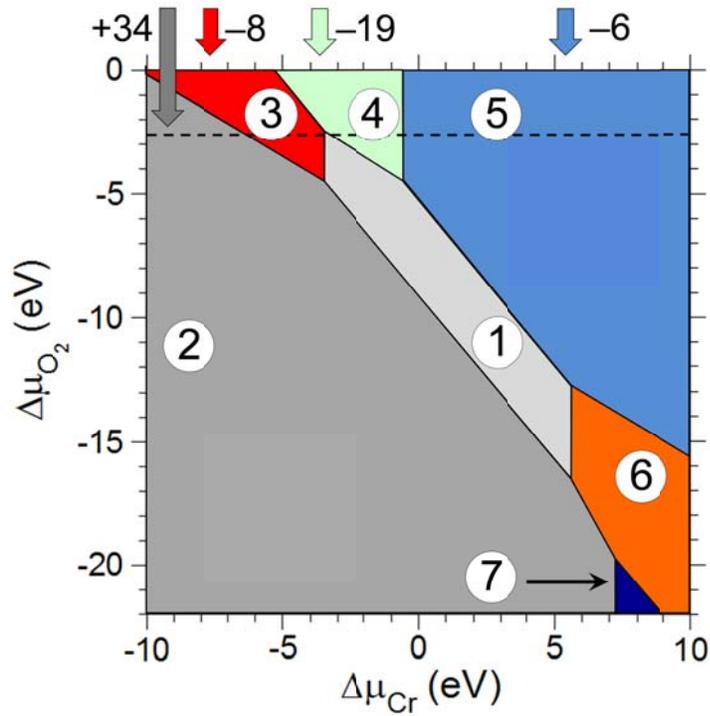

**Figure 5.** (a) Random and channeling RBS spectra obtained using 2.0 MeV He$^+$ ions at a scattering angle of 150$^o$ for LCO films for different *x*. (b) The three spectral windows at which we carried out La angular scans and the depths at which counts in these windows originate ("diffusion" refers to La which diffuses into the STO). (c) to (e) RBS angular scans of the three La-containing film regions along with the Sr scan for STO(001) for stoichiometric (*x* = 1.00), Cr-rich (*x* = 0.85), and La-rich LCO (*x* = 1.10) films, respectively. (f) Cr RBS angular scans for the three LCO films.

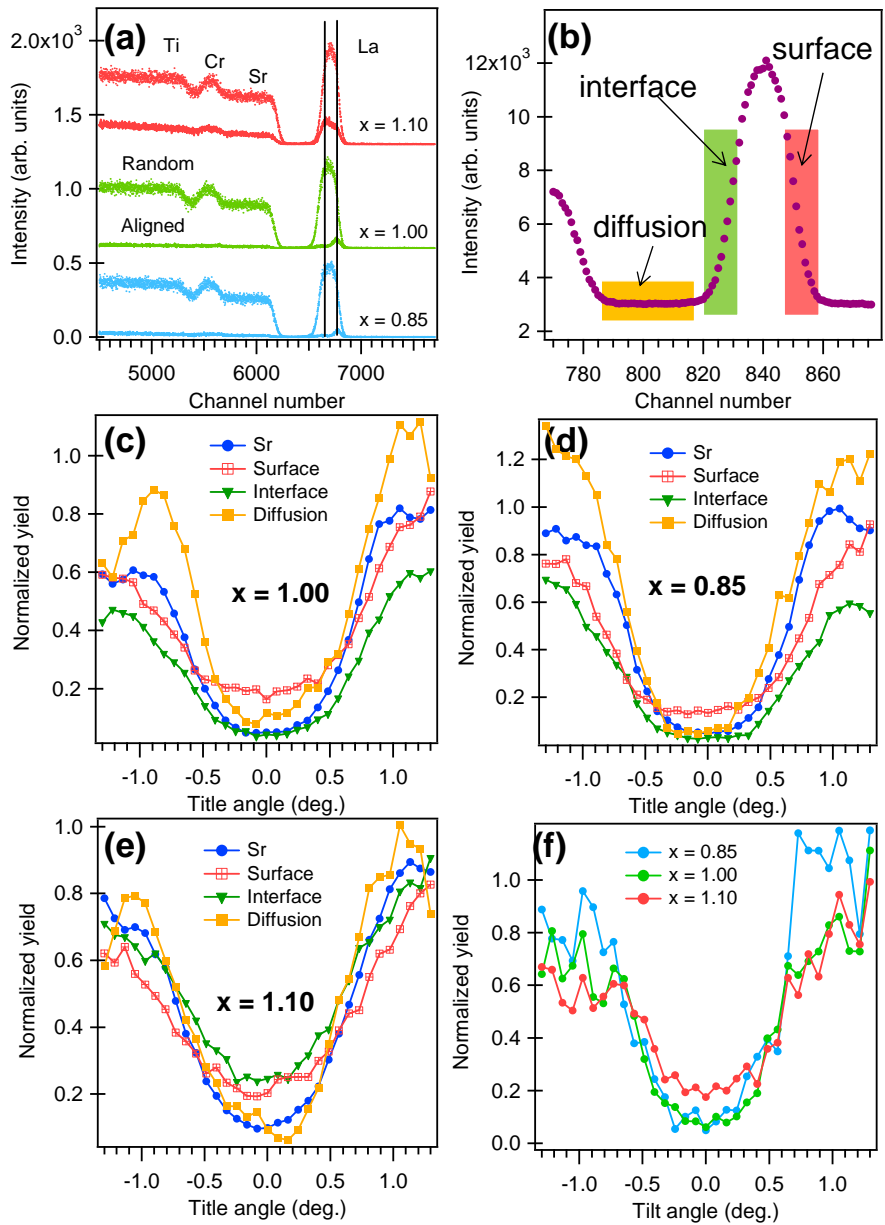

**Figure 6.** (a) RBS spectra in the random geometry (solid circles) and SIMNRA simulations (solid curves) for LCO films with different *x* values. (b) Close-up view of La-Sr valley for each film, along with simulations using both abrupt (dots) and intermixed (solid curves) interface models. (c) The La atom profile used in the intermixed interface simulation for the stoichiometric LCO film.

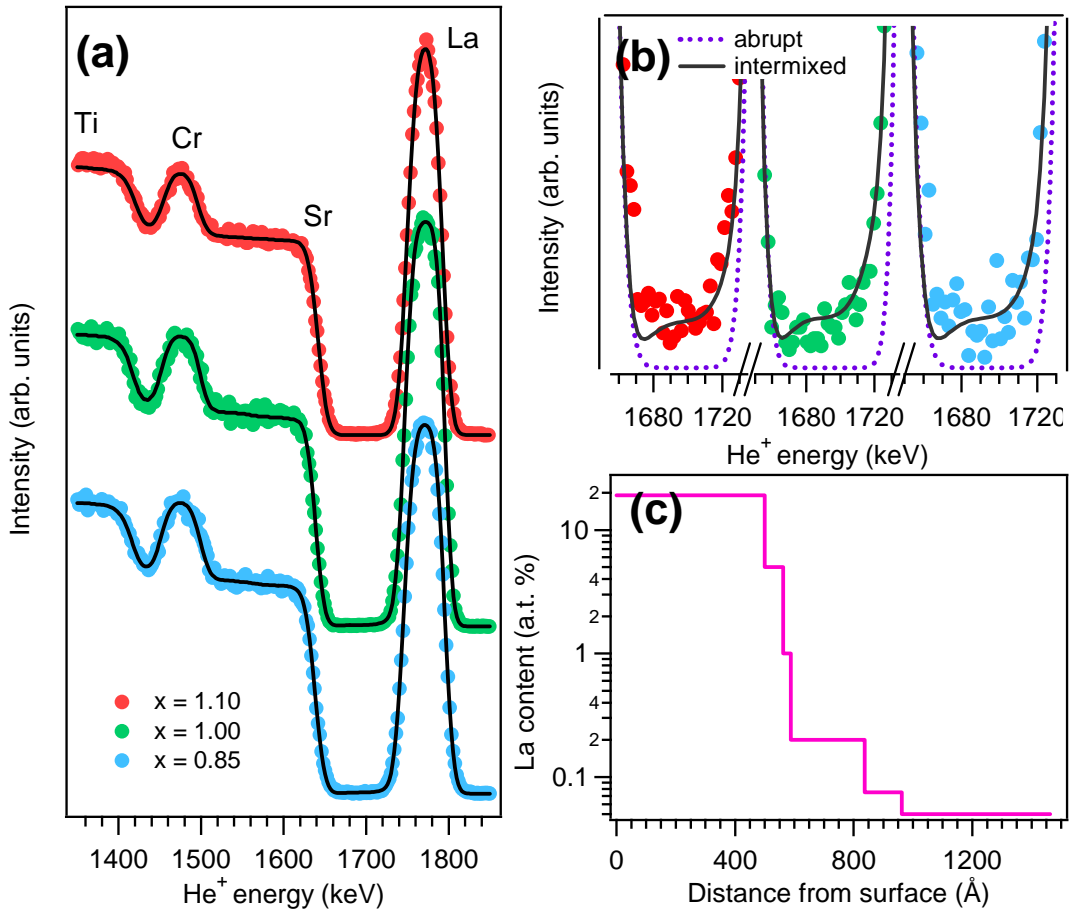

**Figure 7.** (a) RBS spectra in the random geometry (solid circles) for 500Å-thick, MBE-grown epitaxial and amorphous LCO films on STO(001), along with simulations using both abrupt (green curve) and intermixing (blue curve) models of the interface. (b) Close-up view of the La-Sr valley for each film, along with the same simulations. (c) & (d) RBS for a 180Å-thick, MBE-grown LCO film grown on Si(001) at ambient temperature along with an abrupt-interface simulation. (e) & (f) RBS for a 580Å-thick, PLD-grown LAO film grown on Si(001) at ambient temperature, along with both abrupt (green curve) and intermixing (blue curve) interface simulations.

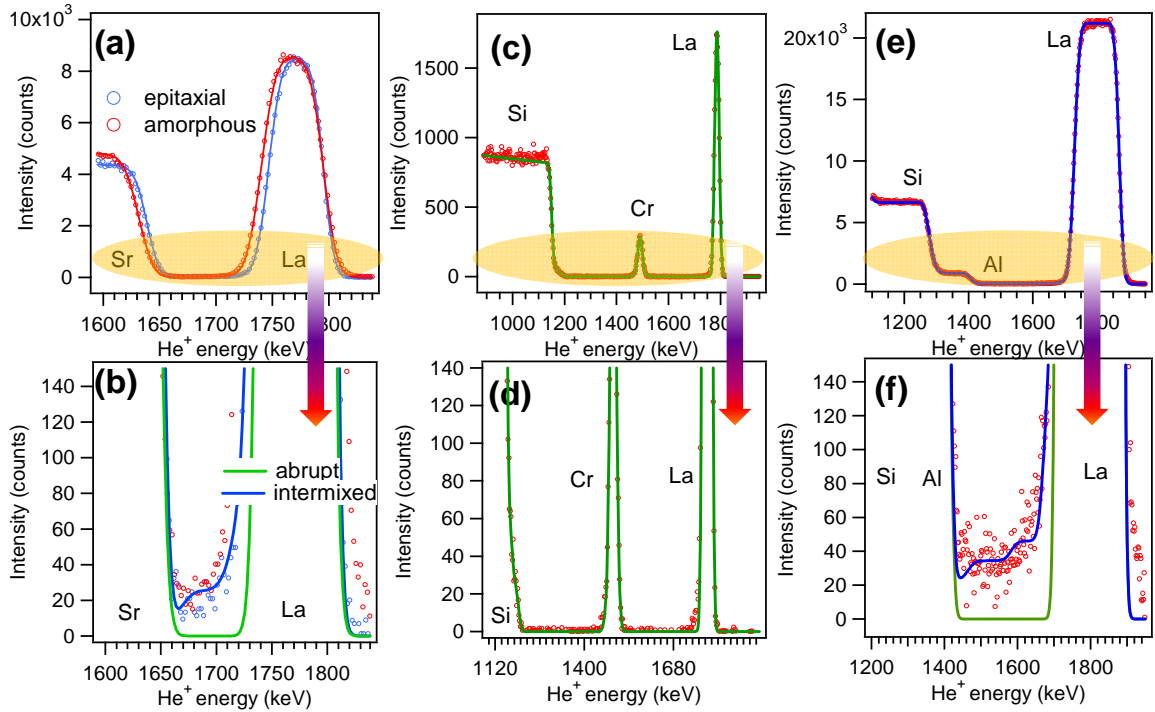

**Figure 8.** Lower- (a & b) and higher-magnification (c & d) STEM images from a single representative region of a stoichiometric 500Å-thick epitaxial LCO film on STO(001), taken in both HAADF-STEM (a & c) and LAADF-STEM (b & d) modes. Both demonstrate a fairly uniform film in the initial ~20 nm, and minor inhomogeneities in the final ~30 nm.

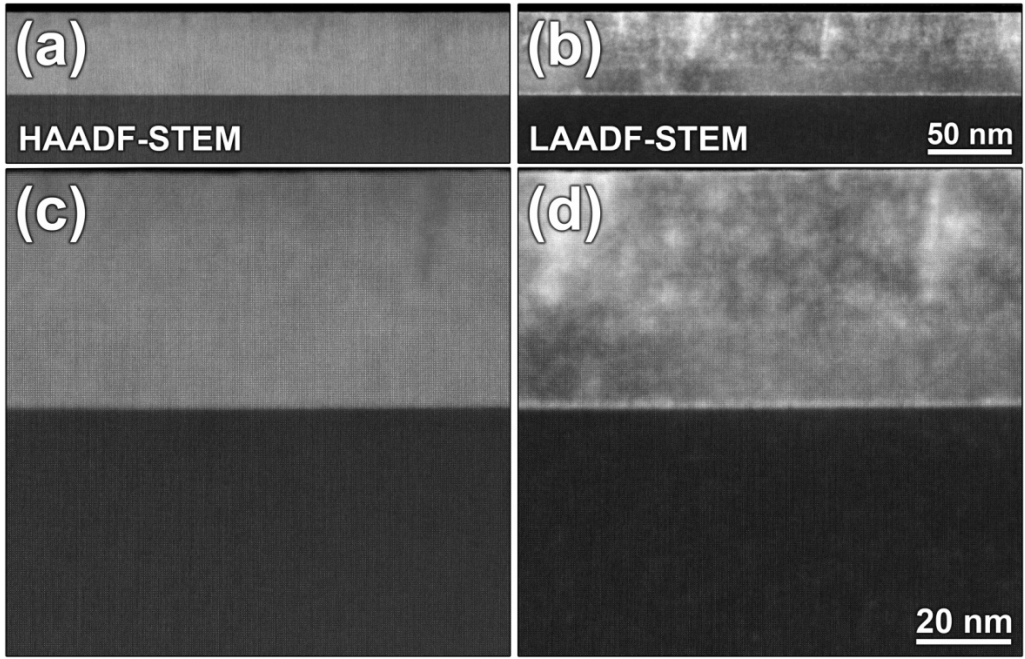

**Figure 9.** HAADF-STEM images of the characteristic defects in the La-rich ~500Å-thick LCO film. (a) At lower magnifications one can see the high density of defects just above the LCO/STO interface, as well as the general in-plane inhomogeneity within the LCO film. (b) These defects vary in size from sub-nm to tens of nm, and would substantially account for the surface roughness and poorer crystal quality observed by other the techniques employed in this work. The regions above these inclusions can be viewed as grain boundaries, punctuated with defects. To illustrate the shift in the (001) planes near these defects, magnified images of the regions immediately above typical defects (c & d) have been severely compressed along a [100] direction (e & f). For typical smaller inclusions (c & e) the boundaries are fairly coherent

stacking faults along [001]. For larger inclusions (d & f) the planes are further bowed, resulting in junctions more like typical low-angle grain boundaries.

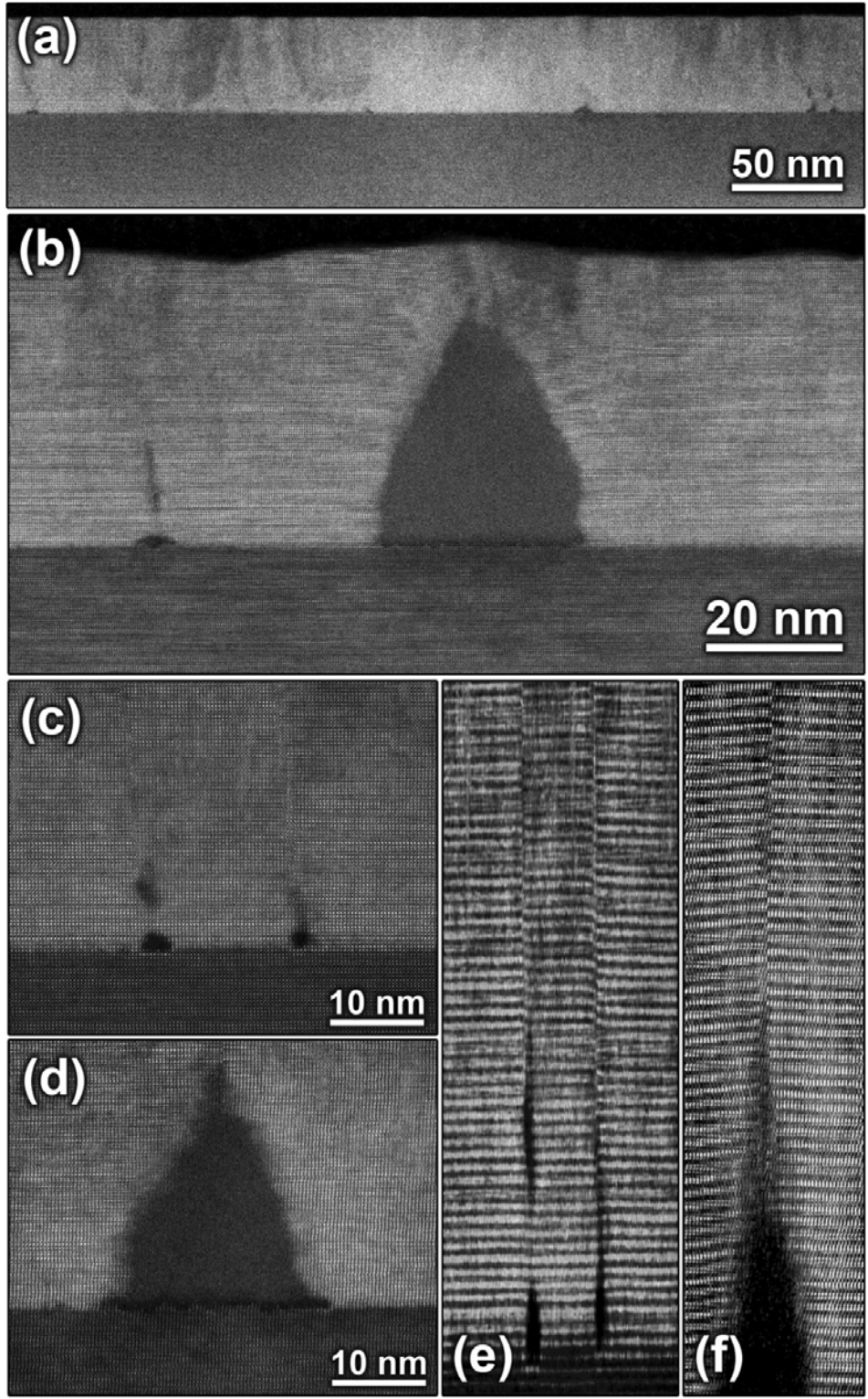